\documentclass[fleqn,usenatbib]{mnras}

\usepackage[T1]{fontenc}
\usepackage{ae,aecompl}

\usepackage{graphicx}	
\usepackage{amsmath}	
\usepackage{amssymb}	

\title[BBH mergers in open clusters] {Binary black holes mergers from hierarchical triples in open clusters}

\author[Britt et al.]{
Dylan Britt $^{1,2}$\thanks{E-mail: djbritt@stanford.edu; erezmichaely@gmail.com},
Ben Johanson $^{1}$,
Logan Wood $^{1,3}$,
M. Coleman Miller $^{1,4}$ 
and Erez Michaely $^{1}$
\\
$^{1}$Department of Astronomy, University of Maryland, College Park, MD 20742, USA\\
$^{2}$Kavli Institute for Particle Astrophysics \& Cosmology, P.O. Box 2450, Stanford University, Stanford, CA 94305, USA\\
$^{3}$Department of Physics and Astronomy,
George Mason University, Fairfax, Virginia 22030, USA\\
$^{4}$Joint Space-Science Institute, University of Maryland, College Park, MD 20742, USA
}

\date{Accepted XXX. Received YYY; in original form ZZZ}

\pubyear{2021}

\begin{document}
\label{firstpage}
\pagerange{\pageref{firstpage}--\pageref{lastpage}}
\maketitle

\begin{abstract}
    A promising channel for producing binary black hole mergers is the Lidov-Kozai orbital resonance in hierarchical triple systems. While this mechanism has been studied in isolation, the distribution of such mergers in time and across star-forming environments is not well characterized. In this work, we explore Lidov-Kozai-induced black hole mergers in open clusters, combining semi-analytic and Monte Carlo methods to calculate merger rates and delay times for eight different population models. We predict a merger rate density of $\sim$1--10\,Gpc$^{-3}$\,yr$^{-1}$ for the Lidov-Kozai channel in the local universe, and all models yield delay-time distributions in which a significant fraction of binary black hole mergers (e.g., $\sim$20\%--50\% in our baseline model) occur during the open cluster phase. Our findings suggest that a substantial fraction of mergers from hierarchical triples occur within star-forming regions in spiral galaxies.
\end{abstract}

\begin{keywords}
    black hole mergers -- celestial mechanics -- galaxies: clusters: general
\end{keywords}



\section{Introduction}

The detection of binary black hole (BBH) mergers via gravitational wave (GW) emission became routine by the O3 observational run of the LIGO and Virgo collaborations. To date, tens of BBH mergers have been detected, with an overall merger rate density of $\mathcal{R}_{\rm BBH}=23.9^{+14.9} _{-8.6}$\rm {Gpc$^{-3}$\,yr$^{-1}$} \citep{Abbott2020b}. The identification of relevant channels which lead to mergers via GW emission is an ongoing endeavour which spans a number of subfields, including orbital dynamics, stellar evolution, and dynamics on the scale of galaxies.  

Channels for BBH mergers may be grouped into four broad categories. 
The first, isolated binary stellar evolution of massive stars \citep[e.g.,][]{Tutukov1973,Tutukov1993,Lipunov1997,Bethe1998,PortegiesZwart1998,Kalogera2000,Mandel2010,Voss2003,Kalogera2007,Belczynski2008,Dominik2012,Dominik2013,Dominik2015,deMink2015,Belczynski2016,Eldridge2017,Giacobbo2018,Olejak2020}, proposes that some massive stellar binaries evolve to short-period binaries prior to either star forming a BH. One type of such evolution occurs during one or two common envelope episodes, in which one star swells during the giant phase, imparting drag on the other and shrinking their mutual orbit. The total orbital energy loss is directly related to the amount of energy transferred to the envelope of the giant star. If the energy transfer is too efficient then the binary merges before the objects turn into BHs; if the transfer is too inefficient then the binary does not lose enough orbital energy to merge via GW emission. The result is a short-period stellar binary which then evolves to a BBH and merges via GW emission within a Hubble time. Studies of this channel predict a delay-time distribution $\propto t^{-1}$ that starts $10-100 \rm{Myr}$ after star formation.  They also predict no measurable eccentricity in the LIGO detection band (due to circularisation during the binary interaction phase and the subsequent circularisation from gravitational radiation) and merger rate densities of $\sim$10$^{-2}$--10$^{3}$\,Gpc$^{-3}$\,yr$^{-1}$. Another isolated binary formation scenario is the chemically homogeneous stellar evolution \citep{Marchant2016,deMink2016,Mandel2016}. In this scenario a massive binary that is close to contact experiences intense internal mixing that keeps the stars chemically homogeneous while the cores are burning hydrogen. The hydrogen in the star is thus nearly exhausted and thus a common envelope phase is avoided. The predicted BBH merger rate is up to $500~{\rm Gpc}^{-3}~{\rm yr}^{-1}$ \citep{deMink2016}.

A second merger channel is dynamical in nature and proposes that existing BBHs are induced to merge in dense environments such as galactic centers, AGN accretion disks, or globular clusters. In these settings, BBHs experience strong gravitational interactions with individual stars or high-multiplicity systems, and these interactions tend to harden the target binaries and may increase their eccentricities  \cite[][]{Sigurdsson1993,Kulkarni1993,PortegiesZwart2000,Madau2001,Miller2002,Gueltekin2004,Gueltekin2006,Miller2009,McKernan2012,Samsing2014,Rodriguez2016,Stone2017,Rodriguez2018,Fragione2018,Banerjee2018,Hamers2018,Leigh2018,Rodriguez2021}. Models of these interactions predict merger rate densities of $\sim$2--25\,Gpc$^{-3}$\,yr$^{-1}$.

The third channel concerns mergers of initially wide, isolated systems, either binaries or triples,  in the field of the host galaxy \citep {Michaely2019,Michaely2020,Michaely2020b}. For wide systems, the field of the host galaxy is considered a collisional environment due to frequent flyby interactions with field stars. These interactions are capable of exciting the eccentricity (in the case of binaries) or outer eccentricity (in case of triples), with the result that mergers occur via increased GW emission (binaries) or three-body instabilities (triples). Predicted BBH merger rate densities for this channel are $\sim$1--100\,Gpc$^{-3}$\,yr$^{-1}$.

The fourth merger channel, and the focus of this paper, is secular evolution in hierarchical triple systems. These systems reside either in the field of the host galaxy \citep[e.g.,][]{Antonini2016,Antonini2017,Silsbee2017} or in dense environments \citep[][]{Miller2002a,Antonini2012,Antonini2014,Kimpson2016,Petrovich2017,Samsing2018,Hoang2018,Fragione2019,Hamilton2019,Martinez2020,Wang2020}. In this channel, a BBH experiences secular effects due to its tertiary companion in the form of the Lidov-Kozai resonance \citep{Lidov1962,Kozai1962,Harrington1968,Lidov1976,Innanen1997,Ford2000,Blaes2002}; for a recent review, see \citet{Naoz2016}. In this resonance, the eccentricity of the BBH experiences cyclic changes which boost the GW emission rate of the inner binary and lead to a merger. Predicted merger rate densities due to this channel are $\sim$0.5--15\,Gpc$^{-3}$\,yr$^{-1}$.

Distinguishing the various channels for producing BBH mergers is important, given that each may yield mergers with particular observational signatures and with different spatial or temporal distributions. BBH mergers in open clusters have been studied previously via $N$-body simulations  \citep{Banerjee2018,Kumamoto2019,DiCarlo2019,DiCarlo2020,Gonzalez2020,Weatherford2021}, which predict merger rate densities of $\sim$0.3\,Gpc$^{-3}$\,yr$^{-1}$ in these environments. \citet{Michaely2018} found that a small fraction, up to fraction of a percent, of mergers are expected to occur extremely close in time to the formation of the second BH, specifically within years to decades following the supernova. Open clusters are loosely bound groups of young stars with stellar number densities $n_* \sim 0.1$--$10$\,pc$^{-3}$ and typical velocity dispersion $\sigma \sim 1$--$5$\,km\,s$^{-1}$ \citep{moraux2016}. We assume that effectively all star formation occurs in these clusters \citep{Lada2003}, which remain bound for lifetimes ranging from $\sim$100$\, \rm Myr$ for the sparsest examples to a few Gyr for the densest clusters \citep{moraux2016}.

In this work, we apply semi-analytic modeling and Monte Carlo simulations to the hierarchical triple channel, studying a set of models describing different initial triple system populations. For each model, we calculate the total BBH merger rate density as well as the cumulative distribution of mergers as a function of time since star formation; this is known as the delay-time distribution (DTD). In particular, we calculate the fraction of mergers which occur while a triple still resides in its birth cluster. The main focus of this work is to estimate the fraction of mergers in the open cluster phase out of the total mergers induced by the secular evolution. 

We begin by describing our semi-analytic treatment of BBH mergers induced by the secular Lidov-Kozai resonance in Section \ref{sec:Kozai}. In Section \ref{sec:Numerics}, we then establish our numerical approach and the different population models considered. Section \ref{sec:Results} presents the simulation results, including the DTD and merger rate for each model. Section \ref{sec:Discussion} discusses our model assumptions and limitations, and in Section \ref{sec:Conclusions}, we summarise and offer broader context for our results.


\section{BBH mergers from hierarchical triples}
\label{sec:Kozai}

In the following section, we briefly describe the secular evolution of triple systems under the Lidov-Kozai resonance. For a more detailed description of this mechanism, see \citet{Naoz2016}.

\subsection{Newtonian treatment}
\begin{figure}
\includegraphics[width=1\columnwidth]{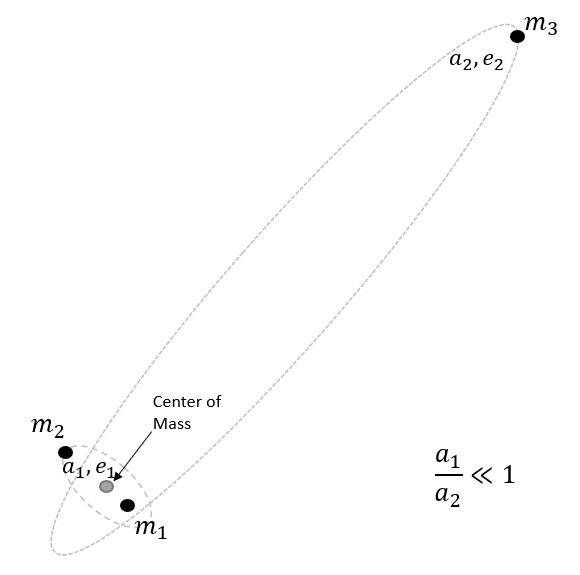}
\caption{Illustration of a hierarchical triple system. The inner binary consists of two objects with masses $m_1$ and $m_2$ whose orbit is defined by a SMA $a_1$ and eccentricity $e_1$. In this study we set $e_1=0$. The outer binary consists of the tertiary of mass $m_3$ and the center of mass of the inner binary. The orbit of the outer binary is defined by a SMA $a_2$ and eccentricity $e_2$. The angle between the planes of the inner and outer binaries is the system inclination $I$.}  
\label{fig:illustration}
\end{figure}

A hierarchical triple system is composed of an inner binary with masses denoted $m_1, m_2$ and a distant tertiary of mass $m_3$. The inner binary is characterised by its orbital semimajor axis (SMA) $a_1$ and eccentricity $e_1$. The center of mass of the inner binary then hierarchically constitutes an additional two-body system with the tertiary; this system is referred to as the outer binary, with SMA $a_2$ and eccentricity $e_2$. Each binary defines a unique orbital plane, and the angle between these two planes is the inclination $I$ associated with the triple system. Within these planes, the orientations of the inner and outer orbits are given by their arguments of pericenter $\omega_1$ and $\omega_2$, respectively. See Fig. \ref{fig:illustration} for a diagram of a general hierarchical triple system.

A three-body system is chaotic when the system masses and separations are similar. Such a system thus tends to break apart on dynamical timescales. Hence, on grounds of system stability, most astrophysical triple systems are hierarchical in scale; i.e.,  $a_1 \ll a_2$. This hierarchy of spatial scales sets a corresponding hierarchy of timescales for these systems: the inner binary orbital period $P_1$ is much shorter than the outer binary orbital period $P_2$, and any dynamical evolution of the system occurs on timescales much longer than both.

When the secular approximation is applied to hierarchical triple systems, one can show that the orbital energies of each binary are conserved quantities, and therefore the SMAs $a_1$ and $a_2$  are constant in time. Long-term changes to the system do occur, however, due to mutual torque and angular momentum transfer between the inner and outer binaries. The result of this secular evolution is simultaneous oscillations of the inner eccentricity $e_1$ and system inclination $I$, such that the total angular momentum of the triple system is conserved; at higher order, the outer orbit can evolve as well. Peak eccentricity in the inner binary occurs at the time of minimum inclination, and vice versa, and these oscillations are referred to as Lidov-Kozai cycles \citep{Lidov1962,koz62}.

Following \citet{Miller2002a} and \citet{VanLandingham2016}, we define a conserved quantity derived from the quadrupole-order Hamiltonian for a hierarchical triple system:
\begin{equation}
\label{W}
   W_{\rm N} = -2\epsilon + \epsilon \cos^2I + 5(1 - \epsilon) \sin^2\omega_1 (\cos^2I - 1)\, ,
\end{equation}
where $\epsilon\equiv 1-e_1^2$. The minimum value of $\epsilon$, which corresponds to the maximum value of $e_1$, occurs when $\omega_1=\pi/2$. Hence, knowing the initial values $\omega_{1,0}$ and $e_{1,0}$, one can exploit the conservation of $W$ to calculate the maximum value of the inner binary eccentricity, denoted $e_{\rm max}$. We note here that the octupole-order result is different \citep{Harrington1968,Ford2000,Blaes2002,Thompson2011,Naoz2013,Michaely2014,Naoz2016} but is beyond the scope of this work. 

\citet{Innanen1997} provide a concise and useful relation between the initial inclination and the maximal eccentricity due to the Lidov-Kozai resonance in the quadrupole approximation when the tertiary dominates the system angular momentum: 
\begin{equation}
\label{e_max_Newton}
    e_{\rm max}=\left( 1-\frac{5}{3}\cos^2 I_0\right)^{1/2}\, ,
\end{equation} 
which implies that for the restricted three-body problem, the inner binary eccentricity tends to unity if $I_0=\pi/2$.
The growth of the inner eccentricity to its maximum value over long timescales is a consequence of coherent perturbations by the potential of the tertiary, specifically inner binary precession. 
If the inner eccentricity is sufficiently high, one might expect the inner binary's components to interact and thus to disrupt this precession. In the following subsection, we consider such an effect in general relativity (GR), namely GR pericenter precession in the inner binary.

\subsection{Post-Newtonian treatment}
In a triple system whose inner binary evolves to high sufficiently high eccentricity, GR precession of the inner binary pericenter becomes nonnegligible. This precession interferes with the coherent perturbations due to the tertiary and suppresses the Lidov-Kozai resonance. Following \citet{Miller2002a}, we account for this quenching effect of GR precession by adding to equation \eqref{W} the following post-Newtonian term:
\begin{equation}
\label{W_PN}
    W_{\rm PN}=\frac{8}{\sqrt{\epsilon}}\frac{M_{1}}{m_{3}}\left(\frac{b_{2}}{a_{1}}\right)^{3}\frac{GM_{1}}{a_{1}c^{2}}\equiv\theta_{{\rm PN}}\epsilon^{-1/2}\, .
\end{equation} 
Here $M_1 \equiv m_1+m_2$ is the total mass of the inner binary, $b_2=a_2\left(1-e_2 ^2\right)^{1/2}$ is the semi-minor axis of the outer binary, $G$ is the Newtonian gravitational constant, and $c$ is the speed of light. Note that we include a term for GR pericenter precession but continue to treat GW emission as negligible for the purposes of the Lidov-Kozai resonance; as a result, the sum of equations \eqref{W} and \eqref{W_PN},
\begin{equation}
\label{W_full}
    W=W_{N}+W_{\rm PN}\, ,
\end{equation} 
remains a conserved quantity. As before, the maximal eccentricity (minimal $\epsilon$) is obtained when $\omega=\pi/2$, and the result in this post-Newtonian treatment becomes
\begin{equation}
\label{e_max_GW}
    \epsilon_{{\rm min}}^{1/2}\approx\frac{1}{6}\left(\theta_{{\rm PN}}+\sqrt{\theta_{{\rm PN}}^{2}+60\cos^{2}I_{0}}\right)\, .
\end{equation} 
This maximal eccentricity can be used to estimate the merger time of the inner binary due to GW emission. The merger timescale for a binary of eccentricity $e\approx 1$ is given by \citet{Pet64} as
\begin{equation}
\label{T_GW}
T_{\rm GW}\approx\frac{768}{425}T_{c}\text{\ensuremath{\left(a_{1}\right)}}\left(1-e^{2}\right)^{7/2}\, ,
\end{equation}
where $T_{c} \equiv a_{1}^{4}/\beta$ is the merger timescale for
a circular binary and $\beta\equiv 64G^{3}m_{1}m_{2}\left(m_{1}+m_{2}\right)/(5c^{5})$. However, in the case of a triple system whose inner binary oscillates between its initial eccentricity $e_{1,0}$  and maximal eccentricity $e_{\rm max}$, the merger timescale due to GW emission is necessarily longer. \citet{Randall2018} analytically estimate the merger time in this case to be
\begin{equation}
\label{merge_time}
T_{\rm merger} = \frac{T_{\rm GW}}{\epsilon_{\rm min}^{1/2}}\, .
\end{equation} 
In this work, we are interested in merger times $T_{\rm merger} < T_{\rm Hubble}$ for the purpose of calculating the merger rate density and DTD of BBH mergers originating from hierarchical triples. In the following section, we describe a method for numerically selecting different triple populations in order to calculate these statistics.


\section{Numerical method}
\label{sec:Numerics}

Our approach to calculating merger rate densities and DTDs is as follows. In each of several population models, described in Sections \ref{subsec:STD} and \ref{subsec:additional}, we employ a Monte Carlo simulation to generate $10^{6}$ representative triple systems. For each triple system in a given model, the model analytically determines whether an inner BBH merger occurs within $T_{\rm Hubble}$ using equation \eqref{merge_time} and records the value of $T_{\rm merger}$ in order to calculate the theoretical DTD. 
 
As stated earlier, we assume that star formation occurs entirely within open clusters \citep{Lada2003} and that as a result, black hole progenitor stars all form simultaneously; i.e., we do not calculate any detailed dynamical effects during the main sequence (MS) phase.

\subsection{Creating a population model}
\label{subsec:popmodel}
Each population model uses a Monte Carlo approach to generate a set of stable, hierarchical triple systems whose inner binaries evolve to BBHs. Although binary stellar evolution processes are beyond the scope of this study, we do consider basic restrictions imposed on triple systems due to their passage through the MS phase. Specifically, we exclude triples whose inner binary components would have interacted as MS stars. We exclude any triple that is considered dynamically unstable by the criterion of \citet{Mardling2001}. Additionally, we work under the simplifying assumption that the probability distributions of all system parameters are independent, meaning that an individual triple system can be generated by drawing each of its parameters independently.

Triple systems are produced in this model by drawing initial stellar masses and orbital parameters, then mapping those stellar masses to final BH masses. To generate the inner binary for a system, we draw the primary mass $m_1$ from the Kroupa initial mass function (IMF) \citep{kroupa2001}, denoted $f_{\rm IMF} \left(m\right)$, with a range $[m_{\rm min},m_{\rm max}]$. Because we are interested in masses of BH progenitors, we concern ourselves only with the upper end of the range of initial masses. The Kroupa and Salpeter IMFs \citep{Salpeter1955} are similar in the high-mass regime, and therefore we do not expect that a different choice of IMF would affect the results presented here. However, the specific choice of IMF is important for the normalisation of the results; see Section \ref{subsec:Normalization}. 

With $m_1$ determined, the next parameter drawn is the inner binary SMA, $a_1$. Motivated by observations \citep{Duchene2013,Moe2016}, we draw the inner SMA from a log-uniform distribution (\"{O}pik's law) over a range $[a_{1,\rm min}, a_{1,\rm max}]$. The mass of the second inner binary object is determined by
\begin{equation}
\label{innerratio}
m_{2}=m_{1}q_{1}\, ,
\end{equation}
where $q_{1}$ is the inner binary mass ratio, drawn from a power law distribution $f(q) \propto q^{\gamma}$. For high-mass stars ($M_* \gtrsim 16\, M_{\odot}$), this distribution covers the range $[0.1,1]$. The power law index $\gamma$ is determined by the SMA of the inner binary, with $\gamma = 0$ for $a_1 <100\, \rm AU$ and $\gamma = -1/2$ for $a_1 > 100\, \rm AU$ \citep{Duchene2013}.

The remaining parameters of the inner binary orbit are its  eccentricity $e_1$ and argument of pericenter $\omega_1$. In order to be conservative with merger time we set the inner binary eccentricity to be zero, $e_1 \rightarrow 0$. Any other choise of inner eccentricity distribution would shorten the merger timescale because of the increase of the maximal eccentricity reached in the Lidov-Kozai resonance \citep{Lidov1976,Naoz2016}. Finally, $\omega_1$ is drawn from a uniform distribution on $[0,2\pi]$. These five parameters define our inner binary progenitor star system.

As mentioned previously, we discard any system whose inner binary would have interacted during the MS phase. To check for such interactions, our method calculates the radii of the progenitor stars and compares these to the stars' respective Roche limits. The stellar radius-mass relation is given by $r_{i} \propto m_{i}^{0.57}$ \citep{demircan1991} and the Roche limit by
\begin{equation}
    R_{1(2)} = a_1\times \frac{0.49q_{1(2)}^{2/3} }{0.6q_{1(2)}^{2/3}+\ln{\left(1+q_{1 (2)}^{1/3}\right)}}\, ,
\end{equation}
where $q_{1(2)}=m_{1(2)}/m_{2(1)}$ \citep{eggleton1983}. A system is discarded if $r_i > R_i$ for either progenitor star, reflecting the likelihood that such a system would have interacted significantly during the MS phase and might have failed to produce a BBH.

We now address the parameters characterising the outer binary. The tertiary mass $m_3$ is set by drawing the outer mass ratio $q_{2}\equiv m_3/M_1$ from a power law distribution $q\propto M_1^\gamma$ with $\gamma = -2$ \citep{Moe2016} over a range $[0.1,1]$. The outer eccentricity $e_2$ is drawn from a thermal distribution $f(e) = 2e$ and the outer SMA $a_2$   from a log-uniform distribution over a range $[a_{2, \rm min},a_{2, \rm max}]$. The final parameter needed to specify the triple system is the mutual orbital inclination $I$; this value is drawn from a distribution function $f\left(I\right)$ which varies by model and is discussed further in the following sections.

With the system parameters fully determined, our method next checks that the triple is indeed dynamically stable. The outer pericenter distance is given by $R^{\rm out}_{P} = a_{2}(1-e_{2})$, and \citet{Mardling2001} define the stability threshold
\begin{equation}
\label{stability}
\kappa = 2.8\left[(1+q_{2})\frac{(1+e_{2})}{(1-e_{2})}\right]^{2/5} a_{1}\, ,
\end{equation}
which specifies the smallest outer pericenter value for which the system remains stable. Accordingly, a system is discarded by our model if $R_P^{\rm out} < \kappa$.

The steps described to this point are sufficient to generate a stable, hierarchical, stellar triple. The initial masses of the three system components must now be mapped to the final masses of the BHs or other objects to which they evolve. When the simulation generates a star of sufficient mass, it converts it into a BH by establishing two mass regimes. For a star whose initial mass $m_i$ falls in the range $20\, M_\odot \leq m_i \leq 60\, M_\odot$, the resulting BH is assigned a final mass $m_i/2$, in keeping with the approximate relation between progenitor mass and final BH mass for stars in this range. A star with initial mass $m_i > 60\,  M_\odot$ is converted to a BH with a final mass of $30\, M_\odot$, reflecting the significant mass loss experienced by very massive MS stars. 

The tertiary is treated differently from the initial binary, as it does not necessarily evolve to a BH. For $m_3 \leq 8\, M_{\odot}$, the simulation checks the MS lifetime for that mass; if it is less than $T_{\rm merger}$, then $m_3$ is converted to a $1$-$M_{\odot}$ white dwarf. In this case, we ignore any expansion of the outer SMA $a_2$, given that the expected mass loss of the tertiary stellar companion in this case is negligible relative to the total mass of the triple. To obtain the total merger time, the original MS lifetime is then added to the merger time for the white dwarf system.

For a tertiary in the range $8\, M_{\odot} < m_3 < 20\, M_{\odot}$, i.e., the mass range for forming a neutron star (NS), the calculation is stopped and the system discarded. In this case, it is expected that the triple system will be disrupted by the natal kick of the NS \citep{Hobbs2005}, precluding any secular evolution.
 
The following section introduces the baseline (``standard'') population model, which adopts the most plausible assumptions for the various triple system parameter distributions. A set of additional models then extends the standard model by modifying a single assumption at a time.

\subsection{Standard model}
\label{subsec:STD}
The baseline model assumes that BHs are formed with no natal kicks, either because of a failed supernova or massive fallback. The limits on the primary mass are set to $m_{1, \rm min} = 30\, M_{\odot}$ and $ m_{1, \rm max} = 100\, M_{\odot}$. While $20$-$M_{\odot}$ O-type stars might produce BHs, there is considerable speculation regarding which mass ranges will yield a natal kick when forming compact objects. By raising the minimum mass for BH formation in our simulation, we impose a conservative buffer which makes it more likely that natal kicks can be neglected. 

This model sets the bounds on the inner binary SMA to $a_{1, \rm min} = 0.1\, \rm AU$ and  $a_{1, \rm max} = 100\, \rm AU$. In keeping with the focus on hierarchical triples, the outer binary SMA is assigned a lower bound of $a_{2, \rm min} = 5a_1$ and an upper bound of  $a_{2, \rm max} = 1000\, \rm AU$. This upper bound is determined by the environment: we do not expect open clusters to contain ultra-wide systems, as these would be ionized due to the relatively high stellar density in the cluster. For a more detailed treatment of the open cluster environment, see Section \ref{sec:Discussion}. As mentioned in section \ref{subsec:popmodel} both the inner SMA, $a_1$, and the outer SMA, $a_2$ distributions are equal in log intervals of $a_1$ and $a_2$ respectively. 

The inclination $I$ of each system is of particular interest when studying the Lidov-Kozai resonance. Given the dearth of observational constraints on the inclinations of high-multiplicity systems within open clusters, we make the reasonable assumption that open clusters and their constituents exhibit a bias toward aligned angular momenta. For triple systems, such a bias favors coplanar orbits. To account for this preference, the standard model draws inclinations from a distribution which increases linearly with $\cos{I}$ in the range $\cos{I} \in \left[-1,1 \right]$; see Fig. \ref{dtd}.

\subsection{Additional models}
\label{subsec:additional}
To probe the sensitivity of merger rates and the DTD to the assumptions used in the standard model, we present several additional models. Each isolates and modifies a single assumption in order to test the robustness of our results.

\subsubsection*{\textbf{No Natal Kicks}}
The standard model excludes primary object masses below $m_{1,\rm min} = 30\, M_{\odot}$ due to uncertainty regarding BH natal kicks below this mass. In this No Natal Kicks model, it is assumed that \emph{all}  BHs are born with no natal kick, and thus $m_{1,\rm min}$ is lowered to the traditionally accepted lower limit of $20\, M_{\odot}$ for BH progenitors. 

\subsubsection*{\textbf{Isotropic Distribution}}
In order to test the sensitivity of our results to the initial distribution of mutual inclinations, this model implements an isotropic (rather than prograde-biased) distribution for $I$. Inclinations are drawn from a uniform distribution of $\cos I\in \left[-1,1\right]$, i.e., from prograde to retrograde mutual inclinations. See Fig. \ref{dtd} for the distribution of inclinations generated by this model.

\subsubsection*{\textbf{Prograde-Only}}
\begin{figure*}
    \includegraphics[width=\textwidth]{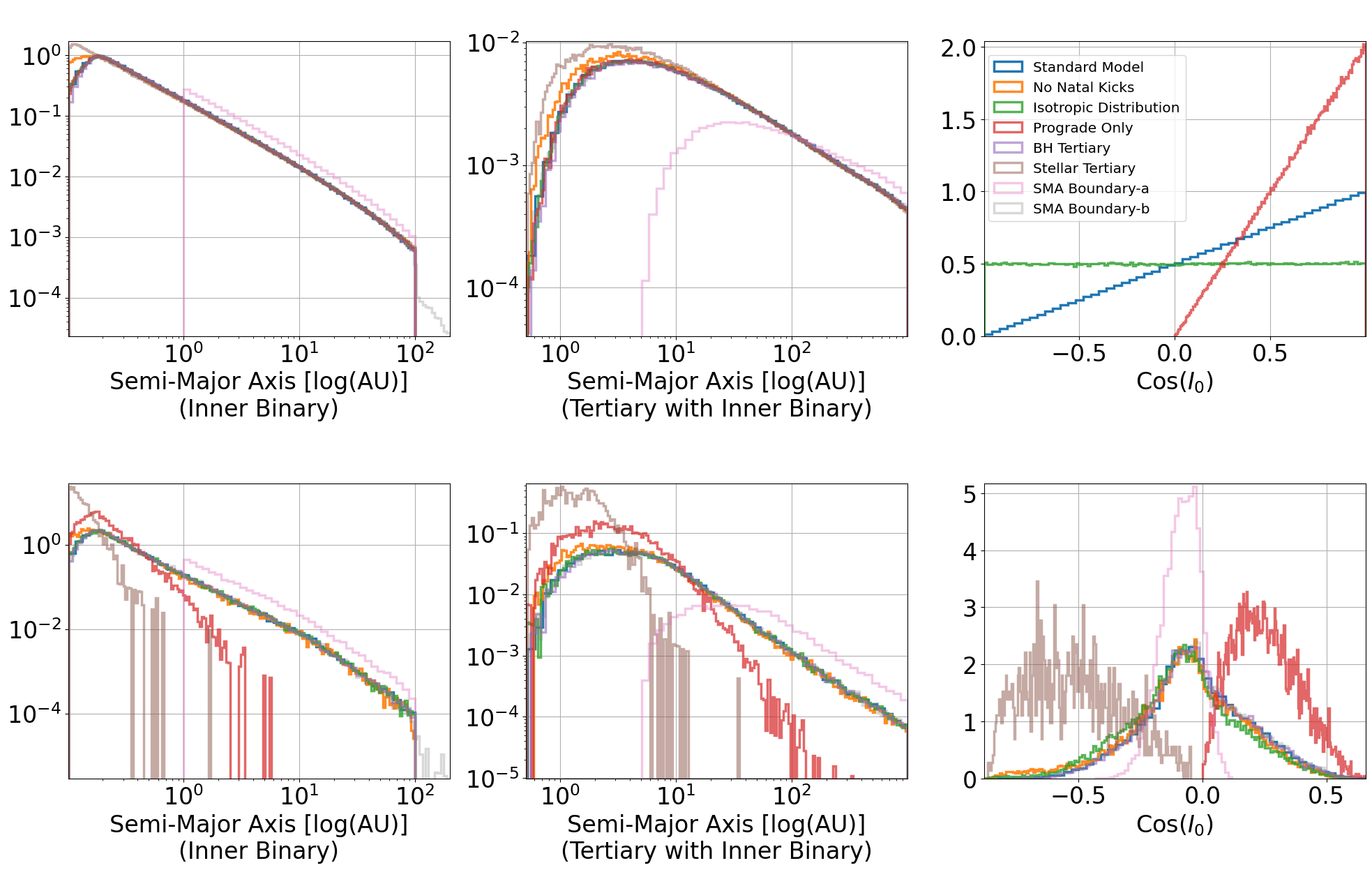}
    \caption{Top row: Initial parameter distributions produced by the Monte Carlo simulation for the various population models. Left panel: inner binary SMA;  middle panel: outer binary SMA; right panel: system inclination plotted as $\cos I_0$. Note that in the distribution of inclinations, only the standard, Isotropic, and Prograde models are shown; all others exhibit no significant differences from the standard model in their distributions of inclinations. Bottom row: the same three parameter distributions shown in the top row, but restricted to the subset of triple systems which merge within a Hubble time in our model. All plots are normalized to unity.}
    \label{dtd}
\end{figure*}

This model restricts the mutual inclination $I$ to prograde values by drawing from a linear distribution of $\cos I\in \left[0,1\right]$. See again Fig. \ref{dtd} for the initial distribution of inclinations.

\subsubsection*{\textbf{BH Tertiary}}
This and the following model concern modifications to the tertiary object in the triple system. In the standard model, the tertiary star is either massive enough to become a BH, forming a hierarchical triple BH, or has a mass low enough to evolve to a white dwarf. Recall that if the tertiary mass falls in the intermediate regime $8\,M_{\odot} < m_3 < 20\, M_{\odot}$, it is assumed to form a NS and disrupt the triple via a high natal kick velocity. In this BH Tertiary population model, only tertiary companions which form black holes are included, and so only systems with tertiary masses $m_3 > 20\, M_{\odot}$ are considered.

\subsubsection*{\textbf{Stellar Tertiary}}
Complementary to the previous model, here only lower-mass tertiary objects are allowed. The evolution of these stars is modeled in two phases, as previously described in Section \ref{subsec:popmodel}. In the first, a star retains its zero-age MS mass $m_3$. In the second phase, the mass of the tertiary star is set to $1\, M_{\odot}$ to account for mass loss during the giant phases and final evolution to a white dwarf. As before, these low-mass tertiaries are restricted to $m_3 \leq 8\, M_{\odot}$.

\subsubsection*{\textbf{SMA Boundaries Model a}}
In all previous models, the inner binary SMA $a_1$ is drawn from the range $\left[0.1\, \rm AU, 100\, AU\right]$. This model considers only larger inner binaries by increasing the lower bound of the inner binary SMA by an order of magnitude, drawing $a_1 \in \left[1\,\rm AU, 100\, \rm AU \right]$.

\subsubsection*{\textbf{SMA Boundaries Model b}}
This model complements the previous model by doubling the upper bound on the inner binary SMA, drawing  $a_1 \in \left[0.1\, \rm AU, 200\, \rm AU\right]$.

\begin{table*}
    \centering
    \caption{Parameter distributions and merger results for the standard and additional models. For comparison with the lifetimes of open clusters, the percentages of systems which have merged at $10^8$ and $10^9$ yr are reported.}
    \label{table_results}
    \begin{tabular}{| p{2.7cm} | p{1.3cm} | p{1.25cm} | p{3.0cm} | p{2.0cm}| p{2.0cm} | p{2.5cm} |}
        \hline
        \textbf{Model} & $\boldsymbol{m_1\, (\rm M_{\odot})}$ & $\boldsymbol{m_3\, (\rm M_{\odot})}$ & \textbf{Inclination} ($\boldsymbol{f(I)}$) & $\boldsymbol{a_1}$\,\textbf{(AU)} & \textbf{Local Rate} & \textbf{Merger Time} \\
        &&&&& \textbf{(Gpc}$\boldsymbol{^{-3}}$\,\textbf{yr}$\boldsymbol{^{-1}}$\textbf{)} & $\boldsymbol{\leq$ $10^9}$ \textbf{yr\, (}$\boldsymbol{10^8}$ \textbf{yr}) \\
        \hline
        Standard & 30--100 & $m\leq8$ & linear in $\cos I$ & 0.1--100  & 6.2 & 49.9 (18.9) \%\\
        && $m\geq30$ \\
        \hline
        No Natal Kicks & 20--100 & $m\leq8$ & linear in $\cos I$ & 0.1--100  & 4.5 & 48.4 (18.1)\%\\
        && $m\geq20$ \\
        \hline
        Isotropic Distribution & 30--100 & $m\leq8$ & uniform in $\cos I$& 0.1--100  & 6.6 & 51.0 (19.5) \%\\
        && $m\geq30$ \\
        \hline
        Prograde-Only & 30--100 & $m\leq8$ & linear in $\cos I$, & 0.1--100  & 2.1 & 33.3 (7.9)\%\\
        && $m\geq30$ & 0 $\leq I \leq$ 1 \\
        \hline
        BH Tertiary  & 30--100 & $m\geq30$ & linear in $\cos I$& 0.1--100  & 6.7 & 50.5 (19.3)\%\\
        \hline
        Stellar Tertiary  & 30--100 & $m\leq8$ & linear in $\cos I$ & 0.1--100  & 0.9 & 15.8 (0.6)\%\\
        \hline
        SMA Boundaries a & 30--100 & $m\leq8$ & linear in $\cos I$ & 1--100  & 3.7 & 51.5 (20.0)\%\\
        && $m\geq30$ \\
        \hline
        SMA Boundaries b & 30--100 & $m\leq8$ & linear in $\cos I$ & 0.1--200  & 6.2 & 33.3 (7.9)\%\\
        && $m\geq30$ \\
        \hline
    \end{tabular}
\end{table*}


\section{Results}
\label{sec:Results}

\subsection{Normalisation and rates}
\label{subsec:Normalization}
We calculate the merger rate density for Lidov-Kozai-assisted BBHs under the assumption that the Milky Way is the prototypical spiral galaxy with a population of $N \approx 10^{10}$ stars. The fraction of primary objects in our triple systems which will form BHs is given by
\begin{equation}
  f_{\rm p} = \frac{ \int_{30\,M_{\odot}}^{100\,M_{\odot}}m^{-2.3}\,dm}{ \int_{0.08\,M_{\odot}}^{100\,M_{\odot}}f_{\rm IMF}(m)\,dm}\ ,
\end{equation}
We continue to treat BHs as forming in high-multiplicity systems \citep{Duchene2013} and without natal kicks. Therefore, taking a uniform distribution of mass ratios $q_1 \in \left[0.1,1\right]$ for the inner binary, the fraction of secondary stars forming BHs is $f_{\rm s} \approx 0.4$. Drawing from a mass ratio distribution $q_2 \propto M_1^{-2}$ for a tertiary at large distances to the inner binary, the fraction of tertiary objects which remain in triple systems is $f_{\rm t} \approx 0.25$. Recall that all tertiary masses in the range $\left[8\, M_{\odot}, 30\, M_{\odot} \right]$ are rejected, as these are expected to disrupt the triple system due to large natal kicks during NS formation \citep{Hobbs2005}. The fraction of the total stellar population which resides in triple systems is taken to be $f_{\rm triple} \approx 0.1$ \citep{Tokovinin2004}. Recall that because this work concerns triples within open clusters, we consider only those triples with an maximum outer binary SMA of $1000\, \rm AU$ and maximum inner binary SMA of $100\, \rm AU$; see Section \ref{sec:Discussion} for a discussion of this choice of values. The fraction of stars which form triple systems with inner binary BBHs that merge via the Lidov-Kozai resonance is then given by
\begin{equation}
F_{\rm model} = 10^{-5} \left(\frac{f_{\rm p}}{10^{-3}}\right) \left(\frac{f_{\rm s}}{0.4}\right) \left(\frac{f_{\rm t}}{0.25}\right) \left(\frac{f_{\rm triple}}{0.1}\right) f_{\rm merger}\, ,
\label{eq:Fmodel}
\end{equation}
where $f_{\rm merger}$ is the merger fraction for hierarchical triples calculated by our numerical model. Recall that a triple system is considered to have merged if $T_{\rm merger} < T_{\rm Hubble}$. The average merger rate for a single Milky Way-like galaxy over a Hubble time is therefore
\begin{equation}
    \Gamma_{\rm MW} = N \times \frac{F_{\rm model}}{T_{\rm Hubble}} \approx 0.53\, \rm {Myr^{-1}}\, .
\end{equation}
Following \citet{Belczynski2016}, the merger rate density in the local universe is given by
\begin{equation}
    \mathcal{R} =\rho_{\rm gal} \times N \times \frac{F_{\rm model}}{T_{\rm Hubble}} \approx 6 \left(\frac{F_{\rm model}}{10^{-5}}\right)  \rm {Gpc^{-3}\, yr^{-1}}\, ,
\end{equation}
where $\rho_{\rm gal} \approx 0.0116\, \rm {Mpc^{-3}}$ is the Milky Way-like galaxy density in the local universe \citep{Belczynski2016}. Depending on the values of the factors that determine $F_{\rm model}$ (see Equation~(\ref{eq:Fmodel}) above), this rate is plausibly comparable to the observed LIGO rate of $\sim$20$\, \rm {Gpc^{-3}\, yr^{-1}}$.

\subsection{Delay-time Distribution}
\begin{figure}
\includegraphics[width=1\columnwidth]{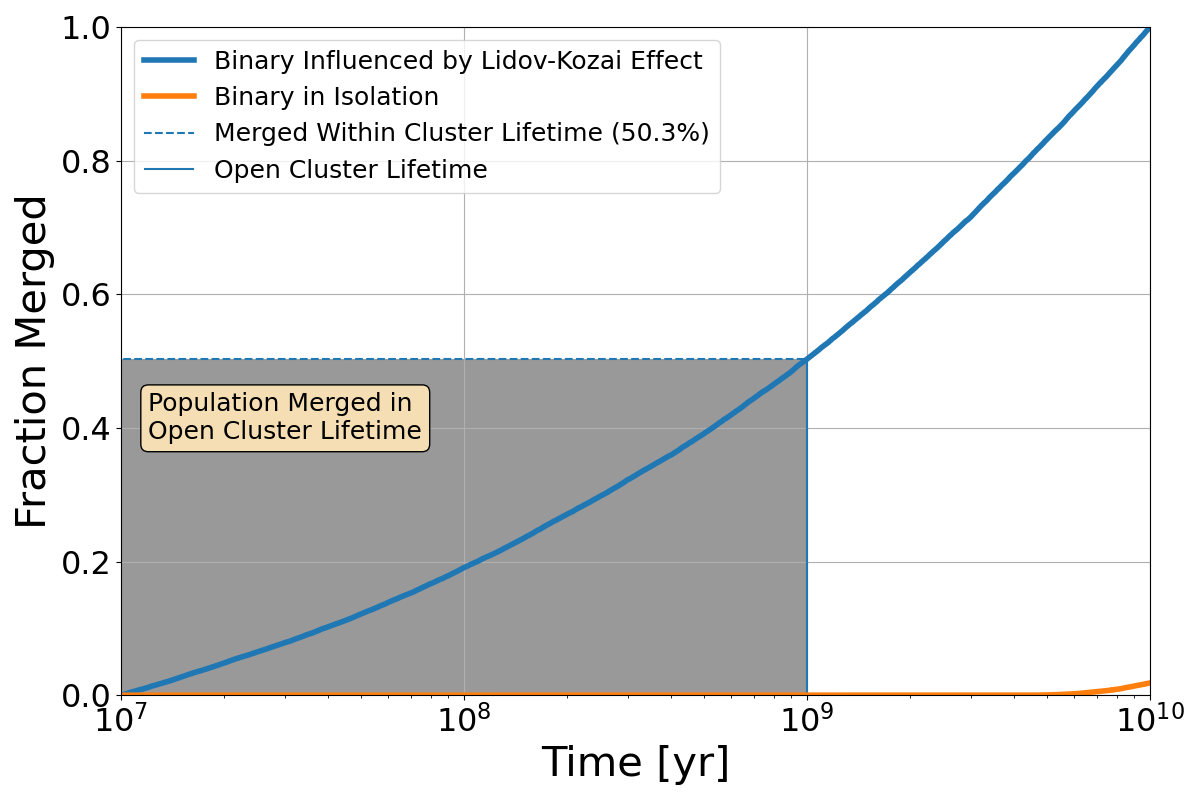}
\caption{Delay-time distribution for simulated mergers in the standard model. The blue curve shows the distribution function for systems which merged within a Hubble time. The gray box indicates the fraction of mergers occurring during the open cluster phase, using $10^9\, \rm yr$ as the upper limit of an open cluster lifetime. For comparison, the orange curve shows the DTD which would result if all BBHs merged in isolation via GW emission alone.}
\label{sm}
\end{figure}

Having recorded $T_{\rm merger}$ for each triple system, we can calculate the fraction of systems which merge within a given time after star formation. Fig. \ref{sm} shows the standard model DTD, i.e., the cumulative merger fraction as a function of time. We find that approximately half of mergers in the standard model occur within the lifetime of open clusters, suggesting that a significant fraction of Lidov-Kozai-induced mergers may occur in these clusters before their dissolution.

\begin{figure}
\includegraphics[width=1\columnwidth]{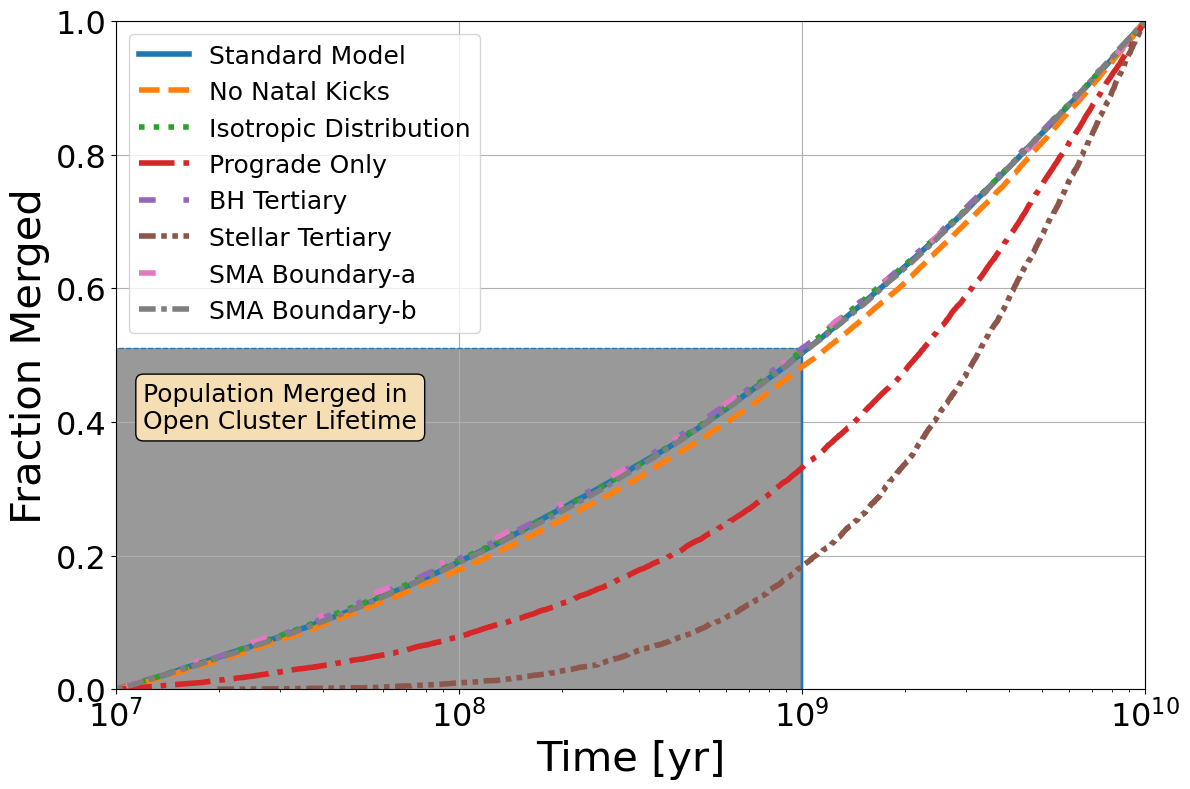}
\caption{Comparison of delay-time distributions for simulated mergers across all population models. Curves show the distribution function for systems which merged within a Hubble time in all models. The gray box indicates the fraction of mergers which occurred during the open cluster phase in the standard model. The distribution shows little sensitivity to initial assumptions, with the exception of low tertiary masses in the Stellar Tertiary model.}
\label{ST}
\end{figure}

Fig. \ref{ST} compares the DTD for the standard model to those for the additional models. Accounting for white dwarf formation in low-mass tertiaries and allowing all viable systems to evolve in time, we find that $\sim$20\%--50\% of Lidov-Kozai-assisted BBH mergers occur within the lifetime of open clusters. We find that the DTD is not particularly sensitive to model assumptions, with the exception of the Stellar Tertiary model, which is skewed toward later merger times and yields a smaller merger fraction within the lifetime of open clusters. This difference can be understood as the result of lower-mass tertiary objects, which have weaker effects on the secular evolution of triple systems.

\section{Discussion}
\label{sec:Discussion}
\subsection{Assumptions}
Each of the population models developed in this work rests on a set of underlying assumptions regarding the parameter distributions of its triple systems. In what follows, we discuss the justification for and implications of several key model assumptions.

\textbf{BH natal kicks}. The first and most important of these assumptions is that BHs are born with little or no natal kick. While it remains unclear whether such kicks are significant \citep{Nelemans1999,Willems2005,Wong2012,Repetto2012,Wong2014,Mandel2016a,Repetto2017}, observational evidence supports BH formation via failed SN or direct collapse \citep{Fryer1999,Ertl2016,Adams2017}. Both mechanisms imply small natal kicks or none at all, supporting the use of our simplifying assumption. In future work, however, we aim to test the importance and sensitivity of this assumption by implementing a more sophisticated population synthesis method.

\textbf{Triple formation}. Throughout this study, we assume that all star formation occurs in open clusters or associations \citep{Lada2003}. The issue of triple formation is not explicitly addressed; our standard model effectively treats each hierarchical triple as primordial. This assumption of primordial system formation is reflected in the non-isotropic distribution of inclinations used in the standard model. We explore a deviation from this assumption by including the Isotropic Distribution model, which draws from a uniform distribution of $\cos I$ and thus simulates triples formed by dynamical processes. In our results, neither the merger rate density nor the DTD depends sensitively on the initial distribution of inclinations.

\textbf{SMA bounds.} In the standard model, the lower bound on the inner binary SMA is set to $a_{1,\rm min} = 0.1$\,AU. For an isolated BBH with $m_1 = m_2 = 20\, M_{\odot}$ and $a_1 = 0.1$\,AU in a circular orbit, equation (\ref{T_GW}) gives an inspiral time (via GW emission only) of $T_{\rm GW}\approx 10^{10}$\,yr, which is on the order of a Hubble time. Therefore, for smaller values of $a_{1,\rm min}$, we would not expect our Lidov-Kozai channel to increase the overall rate of BBH mergers. The upper bound $a_{2,\rm max}$ on the outer binary SMA is set by environmental constraints, specifically the lifetime of a wide orbit in a collisional environment. Following \citet{Bahcall1985} one can calculate the half-life of a wide system of SMA $a_2$ in a collisional environment according to
\begin{equation}
    t_{1/2} = 0.00233\, \frac{v_{\text{enc}}}{G m_{\rm p} n_* a_2}\ ,
    \label{eq:halflife}
\end{equation}
where $v_{\rm enc}$ is the typical encounter velocity at infinity, $m_{\rm p}$ the mass of the perturbing body, and $n_*$ the local stellar number density. For an open cluster, we take $v_{\text{enc}}$ to be a typical velocity dispersion $\sigma\approx 5$\,km\,sec$^{-1}$ and assume a stellar number density $n_*\approx 0.5$\,pc$^{-3}$ and a perturber mass $m_{\rm p}=1\, M_{\odot}$. Taking $10^9$\,yr to be a typical open cluster lifetime, the outer binary SMA of a system whose half-life is equal to the lifetime of the cluster is $a_2 \approx 1000\, \rm AU$; this serves as the upper limit for the size of the outer binary.

\subsection{Mergers in the open cluster phase}
As summarized in Fig. \ref{ST} and in Table \ref{table_results} for all models considered, the fraction of mergers occurring during the lifetime of open clusters is significant. In the standard model, assuming an open cluster lifetime of $10^9$\, Myr ($10^8$\,Myr), we find that $49.9\%$ ($18.9\%$) of BBH mergers induced by the Lidov-Kozai resonance occur in open clusters. This result implies that at least this fraction of mergers from the secular triple channel occur in young environments within star-forming galaxies.


\section{Conclusions}
\label{sec:Conclusions}

In this work, we calculate the merger rates and DTD of BBH mergers occuring in hierarchical triple systems within open clusters via the Lidov-Kozai resonance. This resonance increases the inner binary eccentricity in cycles, allowing the binary to dissipate orbital energy and inspiral via GW emission. Given the sensitive dependence of merger time on orbital eccentricity, BBH mergers in triple systems experiencing the Lidov-Kozai resonance are expected to occur on much shorter timescales than those in isolated binaries. Calculating the DTD for hierarchical triples in open clusters, we find that a significant fraction of mergers ($18\%$--$50\%$ in our baseline model) occur before the open cluster has dissolved. This result suggests that many mergers in hierarchical triples occur in star-forming regions and hence in spiral galaxies.


\section*{Acknowledgements}

E.M. thanks the University of Maryland CTC prize fellowship for supporting this research. The authors thank Selma de Mink, Chris Belczynski and Ilya Mandel for their useful comments on this manuscript.


\section*{Data Availability}

The data underlying this article will be shared on reasonable request to the corresponding author.



\bibliographystyle{mnras}






\bsp	
\label{lastpage}
\end{document}